\documentclass[12pt,preprint]{aastex}

\newcommand{\gtsimeq}{\raisebox{-0.6ex}{$\, \stackrel{\raisebox{-.2ex}{$\textstyle >$}}{\sim}\,$}}
\newcommand{\ltsimeq}{\raisebox{-0.6ex}{$\, \stackrel{\raisebox{-.2ex}{$\textstyle <$}}{\sim}\,$}}          

\shorttitle{Shock-Annealed Silicates}
\shortauthors{Harker and Desch}        

\begin{document} 

\title{Annealing of Silicate Dust by Nebular Shocks at 10 AU}

\author{David E. Harker \altaffilmark{1}}
\affil{NASA Ames Research Center, Space Science Division, Moffett Field,
CA 94035-1000}

\and

\author{Steven J. Desch}
\affil{Dept. of Terrestrial Magnetism, Carnegie Institution of Washington \\
        5241 Broad Branch Rd. NW, Washington DC 20015 }

\altaffiltext{1}{NRC Associate.}

\begin{abstract}
Silicate dust grains in the interstellar medium are known to be mostly
amorphous, yet crystalline silicate grains have been observed in many 
long-period comets and in protoplanetary disks.
Annealing of amorphous silicate grains into crystalline grains requires 
temperatures $\gtsimeq 1000 \, {\rm K}$, but exposure of dust grains in comets 
to such high temperatures is incompatible with the generally low 
temperatures  experienced by comets.  
This has led to the proposal of models in which dust 
grains were thermally processed near the protoSun, then underwent considerable 
radial transport until they reached the gas giant planet region where the 
long-period comets originated.  
We hypothesize instead that silicate dust grains were annealed {\it in situ}, 
by shock waves triggered by gravitational instabilities.  
We assume a shock speed of $5 \, {\rm km} \, {\rm s}^{-1}$, a plausible value
for shocks driven by gravitational instabilities. 
We calculate the peak temperatures of micron and submicron amorphous 
pyroxene grains of chondritic composition under 
conditions typical in protoplanetary disks at 5 -- 10 AU.
Our results also apply to chondritic amorphous olivine grains.
We show that {\it in situ} thermal annealing of submicron and micron-sized 
silicate dust grains can occur, obviating the need for large-scale radial 
transport. 

\end{abstract}


\section{Introduction}

Comets are collections of the dust grains and ices that resided in
the solar nebula.
The dust grains and ices in comets do not appear to have been subject 
to any processing mechanisms since the comets formed at the birth of our 
solar system (Hanner 1999; Brucato et al.\ 1999).
Cometary material therefore is thought to contain pristine solar
nebula material and to represent the physical conditions of the
solar nebula at the birth of the comet, including temperatures 
$\ltsimeq 50 \, {\rm K}$ (Meier \& Owen 1999).
Complicating this interpretation is the spectroscopic evidence of silicate
grains processed at temperatures 
$\gtsimeq 1000 \, {\rm K}$ existing in comets.
Crystalline silicates with distinct mid- and far-IR spectral
resonance features can form a significant component of dust (30 -- 50\%) 
in comet comae (Wooden et al.\ 1999; Harker et al.\ 2002). 
The majority of the crystalline silicate grains seen in many comets must 
have been condensed or annealed (presumably by thermal processing) in the 
solar nebula.
As pointed out by Irvine et al.\ (2000) and Wooden et al.\ (1999),
despite the observations of crystalline olivines 
[$({\rm Mg}_{x}{\rm Fe}_{x-1})_{2}{\rm SiO}_{4}$] and pyroxenes 
[$({\rm Mg}_{x}{\rm Fe}_{x-1}){\rm SiO}_{3})$] ($x$ ranges between 1 
[Mg-pure] and 0 [Fe-Pure]) in circumstellar environments
(Waters \& Waelkens 1998; Malfait et al.\ 1998), crystalline silicates
are essentially unseen in the interstellar medium ($< 5$\%; Li \& Draine 2001).
However, it is not clear how silicate grains could be heated to 
$\sim 1000 \, {\rm K}$ in the cold, comet-forming environment 
($\sim 5 - 100 \, {\rm AU}$). 

In the solar nebula, the high temperatures $\gtsimeq 1000 \, {\rm K}$ 
needed to anneal amorphous interstellar silicate grains, or to condense
crystalline silicate grains directly from the gas phase 
($\gtsimeq 1300 \, {\rm K}$) were achieved in steady-state only 
at distances $\ltsimeq 1 \, {\rm AU}$ from the protoSun at early times
(e.g., Bell et al.\ 1997).
Proposed mechanisms for transporting crystalline silicates to the comet
forming zones include radial diffusion from vigorous thermal convection
(Nuth 2001; Hill et al.\ 2001), or outflows driven by reconnecting magnetic
field lines (``X-winds'' [Shu et al.\ 1996]).
Although X-winds have been invoked as the cause of chondrule formation
(mm-sized spheres of silicate rock found in meteorites [see Jones et al.\ 
2000]), there is compelling evidence (chondrule rim thickness, 
chondrule-matrix complementarity) that chondrules formed {\it in situ}, 
in the asteroid belt region (see summary in Desch \& Connolly 2002; 
hereafter DC02), obviating the need for radial transport. 
The similarities between the annealing of silicate grains and the 
melting of chondrules have been pointed out by Grady et al.\ (2000)
and Nuth (2001).
These similarities between chondrule formation and silicate annealing lead
us to hypothesize that interstellar, amorphous silicate grains were also 
annealed {\it in situ}, at 5 -- 10 AU, by shock waves in the solar nebula. 

In this {\it Letter}, we calculate the thermal histories of submicron 
silicate grains passing through nebular shocks, using the code of DC02
that was used to calculate the thermal histories of chondrules.
We conclude that silicate grains were likely annealed {\it in situ} in
the gas giant planet region by nebular shocks, and we attribute these 
shocks to disk gravitational instabilities.

\section{Annealing in Shocks} 

\subsection{Silicate Grains and their Emissivities}
\label{opacity}

Dust grains of radius $a_{\rm p}$ are heated in shocks by thermal exchange 
with the gas and by frictional gas-drag heating, and cool at a rate 
$4\pi a_{\rm p}^{2} \, Q_{\rm p} \, \sigma T_{\rm p}^{4}$, where $\sigma$
is the Stefan-Boltzmann constant, and $T_{\rm p}$ and $Q_{\rm p}$ are the 
particle's temperature and emissivity. 
We calculate the emissivities of amorphous pyroxene grains of a chondritic
composition [i.e., $({\rm Mg}_{0.5}{\rm Fe}_{0.5}){\rm SiO}_{3}$] 
ranging from 0.1~\micron\ to 1.0~\micron\ in radius, using optical constants 
from Dorschner et al.\ (1995).
Our chosen grain size and composition correspond to grain properties 
determined from the study of IDPs (Bradley et al.\ 1999; Bradley et al.\ 1996),
solar system comets (Harker et al.\ 2002; Mason et al.\ 2001; Schulze, 
Kissel, \& Jessberger 1997), and circumstellar disks (Bouwman et al.\ 2000; 
Malfait et al.\ 1998).
%Furthermore, we find that 
The opacities of amorphous pyroxene and amorphous olivine are similar enough 
($< 5$\% discrepancies) that our results can be extended to
olivine as well.  
The wavelength-dependent emissivity is found as 
a function of particle size using Mie theory. 

The Silicate Evolution Index, or SEI (\S 2.3), which describes the degree
of annealing of silicate material, strictly refers only to Mg-rich silicates,
not the chondritic composition chosen here.
{\it ISO} SWS found Mg-rich ($x > 0.9$) crystalline silicates in comet 
Hale-Bopp and in the circumstellar environments around Herbig Ae/Be stars.
In the experiments of Hallenbeck et al.\ (2000), only 
Mg-rich crystals (forsterite, enstatite) were annealed.
We chose an Fe-rich, chondritic compostion silicate to generate conservative
estimates of the peak temperature of grains.  
Grains heated in shocks by frictional drag heating and thermal exchange with
the gas cool by emitting thermal radiation at $\approx 3 \, \mu{\rm m}$.
The opacity at $3 \, \mu{\rm m}$ is much higher for grains of chondritic 
composition than for pure Mg-rich compositions.
Accordingly, Mg-rich grains cool less efficiently and reach even higher 
temperatures than the grains considered here.  
Mg-rich grains would be more easily annealed than the grains we consider.

\subsection{Calculation of Temperatures} 
\label{code}

To calculate the temperatures of particles overtaken by nebular shocks,
we use the code of DC02, to which the reader is referred for all computational 
details. 
As described in DC02, shocks waves heat, compress, and accelerate gas 
almost instantaneously.
Particles in the gas come into dynamical equilibrium with the gas through
frictional forces, which can heat the particles.
Particles are also kept hot by thermal exchange with the hot gas, and by 
absorbing the radiation emitted by other particles.
Absorption of thermal radiation heats the particles even before they reach
the shock front. 
For the purpose of calculating the radiation field, we assume a dust opacity 
of $4 \, {\rm cm}^{2}$ per gram of gas, due to small ($< 0.1 \, \mu{\rm m}$) 
particles in thermal equilibrium with the gas.
Particles cool by emitting radiation according to their emissivities (\S 2.1).
In practice, the thermal histories of particles and gas are calculated 
assuming a best guess for the radiation field.
An updated radiation field is then calculated, and the thermal histories
recalculated. 
Iterations are continued until the temperatures of particles at each location 
are converged to within $< 10 \, {\rm K}$, which takes on the order of 
$10^{2}$ iterations. 
Since particle temperatures only increase with each iteration, the convergence 
of the solution is well-defined, and the calculated SEI are lower limits. 
All details of the calculation are identical to DC02, with the exception that
gas properties appropriate to the 5 -- 10 AU region are used (\S 2.4), and the 
particle radius is $\sim 1 \, \mu{\rm m}$.
The thermal and dynamical timescales scale with particle radius, so this
small particle radius makes the calculation a factor of $10^{3}$ more 
computationally intensive than the chondrule program.  
Each calculation presented here took a week on a PC with dual 833 MHz
processors, so the SEI is calculated for a small number of parameter
combinations. 

\subsection{Annealing}
\label{annealing}

The annealing of amorphous silicate grains into crystalline grains 
has been studied experimentally by Hallenbeck et al.\ (2000),
who have developed a predictive quantity called the Silicate Evolution 
Index (SEI).  
This quantity is essentially an integration of the temperature $T$ over 
the thermal history of the silicate particle, and is meant to predict
the change in the spectra of silicate particles as they anneal.
At temperatures $\ltsimeq 1067 \, {\rm K}$, the spectra evolve in a two-step
process: as grains spend a time $t$ at high temperatures, their spectra 
evolve, then stall, then continue evolving. 
For temperatures that exceed 1067 K, there is no stall in the evolution of 
the spectra; in this case, 
${\rm SEI} \approx 1 + 5.88 \times 10^{69} \, t \, \exp \left( - 177,430 / T \right)$, with $t$ in seconds. 
Grains with SEI $> 2$ are evolving toward crystalline, and 
SEI $\gtsimeq 100$ are considered completely crystalline.  
Using this formula, we can estimate the peak temperatures grains must reach
to be annealed in the shock.  
In our parameter studies, particles are found to be within 20~K of their peak
temperatures for $\ltsimeq \, 0.1 \, {\rm s}$ (see \S~2.4). 
Values of the SEI exceeding 10 (moderately annealed) therefore require peak 
temperatures $\gtsimeq 1150 \, {\rm K}$. 
Empirically (Table 1), annealing was found to require peak temperatures
$\approx 1200 \, {\rm K}$.

\subsection{Results} 

\begin{figure}[p]
\epsscale{1.0}
\plotone{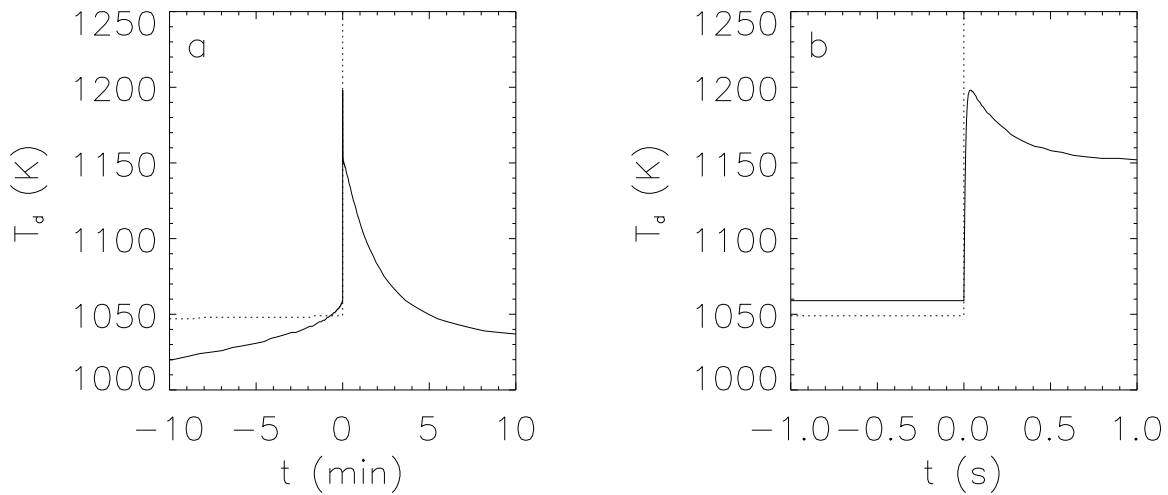}

\caption{Temperatures of a dust particle (solid line) and gas (dashed line)
          in a nebula shock.  The ambient gas density, 
          $5 \times 10^{-10} \, {\rm g} \, {\rm cm}^{-3}$, 
          and shock speed, $5 \, {\rm km} \, {\rm s}^{-1}$, are just 
          sufficient to anneal this $1.0 \, \mu{\rm m}$-radius grain. 
          The temperatures seen by the particle ten minutes before to
          ten minutes after passing through the shock front are shown in 
          (a).  Absorption of thermal radiation emitted by warm dust heats 
          the particle in both the pre-shock and post-shock regions, and 
          thermal exchange with hot gas heats the particle in the post-shock
          region.  A short-lived spike is evident in (b), which displays the 
          temperatures in the two seconds surrounding passage through the
          shock.  Friction with the gas adds to the heating during the time 
          it takes the particle to come into dynamical equilibrium with the 
          gas, and this extra heating to 1200 K leads to annealing of the 
          particle, with ${\rm SEI} \approx 8$.
          \label{fig:temp} }

\end{figure}

The temperature of a $1.0 \, \mu{\rm m}$ grain during the ten minutes
before and ten minutes after being overrun by a  
$5 \, {\rm km} \, {\rm s}^{-1}$ shock is shown in Fig.~\ref{fig:temp}$a$.
As discussed in \S 3.1, we consider this a likely speed of shocks at 
5 -- 10 AU due to gravitational instabilities.
The ambient gas density in this example is 
$\rho_{\rm g} = 5 \times 10^{-10} \, {\rm g} \, {\rm cm}^{-3}$.
Most of the annealing of this pyroxene grain will take place while 
the particle is at its peak temperature, $1200 \, {\rm K}$, at 
which it remains for $\approx 0.1 \, {\rm s}$, as shown in 
Fig.~\ref{fig:temp}$b$.  Integrating over the thermal 
history of this particle yields a value of the SEI $\approx 8$, meaning that 
this particle is marginally annealed by passage of a shock through gas
of this density. 

The peak temperatures of initially amorphous pyroxene grains overrun by
$5 \, {\rm km} \, {\rm s}^{-1}$ shocks, and the corresponding values of the 
SEI, are listed in Table~\ref{tab} for other combinations of gas density and 
particle size.  The particle radii, $0.1, 0.25, 0.5$ and $1 \, \mu{\rm m}$, 
were chosen to match the inferred sizes of crystalline silicate grains in 
comets.  
The ambient gas densities, $0.3, 1, 2, 3, 5$ and 
$10 \times 10^{-10} \, {\rm g} \, {\rm cm}^{-3}$, were chosen to correspond
to the densities computed in the $\alpha$ disk models of Bell et al.\ (1997),
parameterized by the mass accretion rate $\dot{M}$ through the disk and by
the standard Shakura-Sunyaev turbulence parameter $\alpha$.
For example, if $\dot{M} = 10^{-8} \, M_{\odot} \, {\rm yr}^{-1}$
(the median mass accretion rate in T Tauri systems; Gullbring et al.\ 1998), 
and $\alpha = 10^{-4}$, then 
$\rho_{\rm g} = 8.4 \times 10^{-10} \, {\rm g} \, {\rm cm}^{-3}$ 
at 5 AU, and $3.0 \times 10^{-10} \, {\rm g} \, {\rm cm}^{-3}$ at 10 AU.
For $\alpha = 10^{-2}$, the values are 
$2.8 \times 10^{-11} \, {\rm g} \, {\rm cm}^{-3}$ at 5 AU and 
$1.2 \times 10^{-11} \, {\rm g} \, {\rm cm}^{-3}$ at 10 AU. 
The post-shock gas densities are greater than the ambient density by 
a constant factor, $\approx \, 6$, and the post-shock temperature are 
set largely by the shock speed only. 
For many of the cases seen in Table~\ref{tab}, we find SEI $\gtsimeq 1$.
Therefore, shock-annealing of silicate grains in the 5 -- 10~AU region is 
evidently possible. 

\section{Discussion} 

Annealing of silicate grains can be accomplished by shocks in the outer 
solar nebula (at 10 AU).
Smaller grains are more easily annealed due to their poorer radiative 
efficiencies.  These results are not affected by the presence of ice mantles, 
because such mantles would sublimate before reaching the shock front.
Water ice is at least as good an absorber as silicates of the thermal 
radiation emanating from the shock front, and dust grains can easily absorb 
enough energy from the radiation field to heat above 150 K.
Figure~1$a$ shows that a dust grain attains temperatures of $> 1000$~K for
$> 10$~minutes. Since the grain is in radiative equilibrium,
the grain is absorbing and emitting thermal radiation at a rate of 
$4\pi a_{\rm p}^{2} \, Q_{\rm p} \, \sigma T_{\rm p}^{4}$.  If the grain
is assumed to be made entirely of water ice with latent heat of sublimation
$2.83 \times 10^{10}$~erg~g$^{-1}$ and radiative absorption efficiency
$Q = 0.1$, it can easily be shown that the ice will completely evaporate
in $< 0.2$~s.  This evaporation time is much less than the 10~minutes the
particle stays at temperatures $> 1000$~K.  Shocks raise grains to high 
temperatures whether or not they have ice mantles.
Therefore, provided shock speeds $\sim 5 \, {\rm km} \, {\rm s}^{-1}$ can be 
attained, annealing of silicates {\it in situ} is very likely. 

Shocks are very likely to occur early in the evolution of the solar nebula.
The accretion disk models of Bell et al.\ (1997) predict
that T Tauri disks with mass accretion rates 
$10^{-8} \, M_{\odot} \, {\rm yr}^{-1}$ and $\alpha = 10^{-2}$ are 
locally gravitationally unstable beyond 11 AU, and disks with 
$\alpha = 10^{-4}$ are locally unstable beyond 6 AU.
Since mass accretion rates $\sim 10^{-8} \, M_{\odot} \, {\rm yr}^{-1}$
are likely to persist for many Myr in T Tauri systems (Gullbring
et al.\ 1998), gravitational instabilities are likely to persist 
in the outer regions of protoplanetary disks.
Since the gravitational instabilities manifest themselves as global
modes (Laughlin \& Rozyczka 1996), they are likely to affect the bulk 
of the disk, even where it is locally stable.

How the gravitational instabilities in a marginally unstable
disk manifest themselves has been explored numerically by Boss (2000, 2001a,b),
who discovered the development of gravitationally bound clumps and
highly non-axisymmetric density structures associated with the clumps
in such disks. 
These structures would be
overdense by orders of magnitude, oriented more or less 
radially, extending inward from the clumps and co-rotating with them
(see Figure 1 of Boss [2000]).
Gas orbiting closer to the Sun would orbit faster than these density
structures and collide more or less normally with them, at relative
velocities comparable to the difference in Keplerian angular velocities 
between the orbiting gas and the clump driving the structure.
For example, a clump at 10 AU would drive a density structure with an
orbital period $\approx 30 \, {\rm yr}$; material orbiting every
$15 \, {\rm yr}$ at 6 AU would slam into this structure at a relative 
velocity $\sim 6 \, {\rm km} \, {\rm s}^{-1}$.  

If nebular shocks existed in the outer solar nebula, whatever their source, 
we find that submicron silicate grains could readily be annealed by 
$5 \, {\rm km} \, {\rm s}^{-1}$ shocks, if the gas 
densities were $\gtsimeq 10^{-10} \, {\rm g} \, {\rm cm}^{-3}$. 
These densities are plausible for the 5 -- 10~AU region of the solar nebula,
according to the $\alpha$ disk models of Bell et al.\ (1997),
but not beyond 10 AU.
Therefore, comets forming in the Kuiper Belt Region (greater than 35~AU) should
not contain crystalline silicate dust annealed by shocks.
Radial diffusion by turbulence is unlikely to change this result. 

Within the time a comet takes to form, $t \sim 10^{5} \, {\rm yr}$ 
(Weidenschilling 1997), turbulent diffusion mixes gas across distances 
$\sim (\alpha \, C H t)^{1/2}$, where $C$ is the sound speed and $H$ the scale 
height of the nebula.
For $\alpha \approx 10^{-3}$, the mixing distance over $10^{5}$ years at 
10 AU is $\ltsimeq 5 \, {\rm AU}$; over $10^{6}$ years it is 
$\ltsimeq 15 \, {\rm AU}$.

The vast majority of short-period
comets, which come from the Edgeworth-Kuiper belt and which
mostly formed beyond 35 AU, should contain amorphous grains only.
The few short-period comets with crystalline silicates must have 
been scattered to the Edgeworth-Kuiper belt after forming at 5 -- 10~AU, 
which is possible dynamically (Duncan \& Levison 1997).
However, most comets formed at 5 -- 10~AU will instead be scattered into
the Oort cloud and be observed entering the solar system today as long-period
comets. 
It is in long-period comets that crystallinity should be expected.

The observational data confirm the prediction that crystalline
silicates should be found almost exclusively in long-period comets.
Fits to Hale-Bopp's (C/1995 O1) 11.2~\micron\ features have 
led to the determinations that, even pre-perihelion, 
$\approx 20 \%$ of all silicates (Hanner et al. 1997, Wooden et al.\ 1999), 
and $\approx 30 \%$ of the olivine silicates (Harker et al.\ 2002) must 
be in crystalline form.
According to Yanamadra-Fisher \& Hanner (1999), comets with comparable peaks
at 10 and 11.2~\micron, such as P/Halley, Bradfield 1987 XXIX, Mueller, 
Levy 1990 XX, and Hale-Bopp, are well fit by a combination of 20\% 
crystalline and 80\% amorphous silicates. 
The crystalline silicate feature at $11.2 \, \mu{\rm m}$ has also been
observed in Hyakutake (C/1996 B2; Mason et al.\ 1998), and 
McNaught-Hartley (C/1999 T1; Lynch, Russell \& Sitko 2001).
Not all long-period comets have crystalline silicates ---Hanner et 
al.\ (1994) found no evidence for annealing in 
Kohoutek 1973 XII, Austin 1990 V, Okazaki-Levy-Rudenko 1989 XIX, and Wilson 
1987 VII ---but most do.
On the other hand, only in one short-period comet, 103/P Hartley 2, have 
crystalline silicates been observed (Crovisier et al.\ 1999).
Data are inconclusive or negative in others (Harker et al.\ 1999;
Hanner et al.\ 1996), including comets P/Brorsen-Metcalf 1989 X 
(Hanner et al.\ 1994), P/Churyumov-Gerasimenko (Hanner et al.\ 1985), 
P/Grigg-Skjellerup (Hanner et al.\ 1984), and P/Schaumasse 
(Hanner et al.\ 1996), each indistinguishable from blackbodies (Hanner 
et al.\ 1994).  Winds hypothetically carrying annealed grains from the inner 
solar system would preferentially seed Oort-cloud comets with crystalline 
silicates, but these winds could possibly also deposit crystalline silicates 
in Kuiper-belt comets as well. 
The low gas densities and orbital velocities of the shock model rule out this
possibility for comets that form beyond $\sim$ 10 AU. 

If shocks annealed silicate grains {\it in situ},
the possibility exists that the nebula gas could be chemically processed 
by the shock as well.  For example, Kress et al.\ (2001) suggest that 
nebula shocks could produce nitriles that are incorporated into 
nitrogen-bearing organics in meteorites.  Observed correlations of 
shock-produced, disequilibrium chemicals like nitriles, with silicate 
crystallinity, would strengthen the case for shock annealing.

Comets, nearly all of them long-period comets, despite forming at very cold
temperatures, contain crystalline silicate grains that imply temperatures
$\gtsimeq 1100 \, {\rm K}$, if only for a fraction of a second.
These conditions are met in nebular shocks. 
Gravitational instabilities make shocks a likely event in the first few Myr 
of the solar nebula, but whatever the source, silicate grains can be annealed
by shocks in the 5 -- 10 AU region of the solar nebula. 
Shock-annealing obviates the need for large-scale radial transport of annealed 
grains from the inner solar nebula to the 5 -- 10 AU region where the 
Oort-cloud comets formed. 

\acknowledgements
D.E.H.\ gratefully acknowledges support from a NRC Fellowship and the 
contributions from Diane Wooden, Chick Woodward and K.\ Robbins Bell.  
S.J.D. gratefully acknowledges support from a Carnegie Fellowship, and 
thanks Alan Boss for useful discussions of nebular shocks.

\begin{deluxetable}{lcccc}
\tablewidth{0pc}
\tablecaption{Peak Temperature and SEI of Annealed Grains \label{tab}}

\tablehead{
 & \multicolumn{4}{c}{\underline{grain radius}} \\
\colhead{$\rho$ (g cm$^{-3}$)} & \colhead{0.1~\micron} & 
\colhead{0.25~\micron} & \colhead{0.5~\micron} & \colhead{1.0~\micron}
}

\startdata
$0.3 \times 10^{-10}$ & 1100~K & \nodata & \nodata & 670~K \\
 & $4.0\times10^{-6}$ & & & $4.9\times10^{-9}$ \\
$1.0 \times 10^{-10}$ & 1305~K & 1095~K & 990~K & 920~K \\
 & $6.6\times10^{4}$ & $2.2\times10^{-6}$ & $3.6\times10^{-9}$ & $1.1\times10^{-8}$ \\
$2.0 \times 10^{-10}$ & 1405~K & 1210~K & 1110~K & 1035~K \\
 & $3.3\times10^{ 9}$ & $4.1\times10^{0}$ & $7.3\times10^{-6}$ & $1.1\times10^{-8}$ \\
$3.0 \times 10^{-10}$ & \nodata & \nodata & 1175~K & 1110~K \\
 & & & $4.3\times10^{-2}$ & $2.7\times10^{-6}$ \\
$5.0 \times 10^{-10}$ & \nodata & \nodata & 1255~K & 1200~K \\
 & & & $3.8\times10^{2}$ & $8.3\times10^{0}$ \\
$10.0 \times 10^{-10}$ & 1650~K & \nodata & \nodata & 1300~K \\
 & $2.4\times10^{16}$ & & & $6.0\times10^{3}$ \\

\enddata

\end{deluxetable}

\newpage

\end{document}